\documentclass[10pt,twocolumn]{IEEEtran}

\usepackage{graphicx}
\usepackage{moreverb} 

\usepackage[cmex10]{amsmath}
\usepackage{algorithmicx}
\usepackage{epsfig}
\usepackage{subfigure}
\usepackage[linesnumbered,lined,ruled]{algorithm2e}
\usepackage{xcolor}
\usepackage{array}
\usepackage{cite}
\usepackage{amssymb}

\begin{document}
\title{\textcolor{blue}{Internet of Radars (IoR): Internet of RAdio Detectors And Rangers}}

\large\author{
\centering \vspace{.2cm}{\"{O}zg\"{u}r B. Akan} \emph{Fellow, IEEE} and Muharrem Ar\i k \emph{Student Member, IEEE}\\
\thanks{M. Ar\i k and \"{O}. B. Akan are with the Next generation Wireless Communications Laboratory (NWCL) Department of Electrical \& Electronics Engineering in Koc University, Istanbul, Turkey. Email: {marik@ku.edu.tr, akan@ku.edu.tr}
}}

\maketitle
\normalsize



\begin {abstract}

\textbf{The Internet of Things (IoT) are interconnected devices for exchanging information through sensors and actuators. One of the main physical sensors to understand the environment beyond the visible world is a radar. Basically, radars have always been a military tool to be used to investigate the environment. However, with the developing technology, radars have become more compact and affordable to use in a building, in a car, in a drone or even in a wristwatch. In the near future, radar-equipped IoT platforms will start to appear increasingly. For each IoT platform, dual use of spectrum with dual aperture is required for sensing and communicating when using conventional approaches. For the radar sensing IoT devices, the emission from the radar and communication circuitry is the main reason of the increase in energy consumption.  Furthermore, an increasing number of radars emerges as congested spectrum, and RF convergence between radars and communication systems becomes more likely to present itself. Recent years, there have been numerous researches which propose using the one emission/waveform for perceiving the environment and sending information. They are often called as ``Joint Radar-Communication (JRC)" systems. Due to the latest developments in JRC system designs, radar sensing IoT platforms now can be transformed into an ``Internet of RAdio Detectors And Rangers (\textcolor{blue}{IoR})". In this article, we present a \textcolor{blue}{short} survey on JRC technologies, possible application areas IoR applications and challenges and future research directions for enabling the concept of \textcolor{blue}{IoR}.}

\end {abstract}

\section{\textcolor{blue}{Introduction}}
\label{sensvssend}
Researchers in radar and wireless communication domain always look for different motivations. One develops techniques for better target parameter estimation when coping with noise, clutter or interference, the other creates mechanisms on achieving maximum capacity under noisy channel. 

JRC system gives an important opportunity to reduce spectrum usage, minimize energy consumption and product cost while doing concurrent operation, as target sensing via radar processing and establishing communication links. The main research area for JRC systems has begun to concentrate around optimization techniques. Researchers focus on finding an optimized way of sending information to corresponding receiver while sensing the environment. Due to its heterogeneous topology, IoT habitat may present great challenges on deciding priority on sensing versus sending. Using JRC technology in IoT structure has led to grow a whole new trade-off environment where both sides should begin to agree.

\textcolor{blue}{Developing IoT technology causes a significant increase on the number of communicating devices. Besides, after recent developments on system on chip structures, more and more small and compact radars are now available at the markets. Increasing number of RF devices emerges as congested spectrum, and RF convergence between radars and communication systems becomes more likely to present itself.}

JRC capable IoT structure may involve large number of dynamic nodes, such as the Internet of Vehicles (IoV). This necessitates building a new platform called, ``Internet of RAdio Detectors And Rangers (RADARs)", IoR, which provides all nodes are effective and connected.

\section{Joint Radar-Communication \textcolor{blue}{Technologies}}

\textcolor{blue}{A JRC system can propose cost-effective solution with concurrent operation, as target sensing via radar processing and establishing communication links. We have divided JRC technologies into four groups in terms of utilized waveform types, i.e. Coexistence, Co-Design, Rad-Com and Com-Rad.}

\subsection{\textcolor{blue}{Coexistence}}

\textcolor{blue}{Researches try to find new techniques to solve radar and communication interference mitigation and spectrum management solutions, i.e. opportunistic spectrum access, interference channel estimation and optimal receiver designs \cite{surveyJRC}. Mainly, these efforts spent for the utilization of communication systems. Current radar sensing IoT application areas do not overlap each other. Also, power of the RF emission is limited for most of the IoT devices. However, after 5G NR bands standardization efforts, radar and communication bands may overlap each other in the near future and coexistence solutions may become feasible to implement in the IoR.}

\subsection{\textcolor{blue}{Co-Design}}

\textcolor{blue}{One of the important promising system solution for RF convergence problem is merging radar and communication capabilities via single waveform, i.e. dual-function radar-communication (DFRC). Co-Design solutions aim to find a suitable unique waveform set to perform best operation for DFRC systems.} In \cite{FrFTRadarWaveform}, data is embedded into chirp sub-carriers of the multi-carrier waveform using Fractional Fourier Transform (FrFT). This method is validated experimentally.

\subsection{Communication Utilized by Radar Systems\textcolor{blue}{: Rad-Com}}

Several techniques are proposed over the years for different types of radar systems with different topologies \cite{surveyRFCommRadar}. Almost all types of radars and all types of modulation are the subject to this research domain \cite{surveyJRC,TowardmmWaveJRC}. Mostly, efforts are concentrated \textcolor{blue}{on improvements in} data-rate, modulation types and transparency over radar operation, \textcolor{blue}{then the} proposed techniques are \textcolor{blue}{grouped} into two.

\subsubsection{Information Embedded Waveform Diversity} The oldest technique is embedding information into radar chirps. In these methods, the information can be transmitted to the direction of interest via emitting one waveform during each radar pulse from a group of predetermined waveforms. Despite the simplicity of the implementation of the method, this modulation scheme degrades target visibility due to the range sidelobe modulation (RSM) caused by varying waveform during coherent processing interval (CPI). To reduce RSM, \textcolor{blue}{several} information embedding approaches are introduced in literature \cite{surveyJRC}. Further, various intra-pulse modulation, i.e. Binary Phase Shift Keying (BPSK) and Minimum Shift Keying (MSK), schemes to carry information symbols are developed for frequency modulated radar waveforms \cite{surveyJRC}. In these solutions, the main goal is not disturbing the operation of the radar.

\subsubsection{Symbol Construction via Transmit Beamforming} 
\begin{figure}[t]
	\centering
	\includegraphics[width=1\columnwidth]{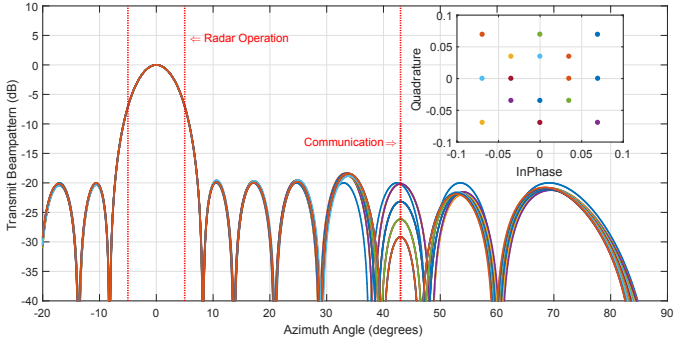}
	\caption{16-QAM transmit beampattern \textcolor{blue}{is given} for a DFRC system with $16$ element transmit antenna array. Communication is realized at $43^{\circ}$ azimuth angle when radar operation is occurred around $0^{\circ}$.}
	\label{fig:DFRC}
\end{figure}

The number of phased array radars in the operation field is significantly increased due to their numerous advantages, \textcolor{blue}{as} digital beamforming opportunity at the receiver and flexible antenna beampattern. Manipulating parameters of each antenna on pulse to pulse basis presents an important opportunity for transmitting information towards direction of interests. Using this \textcolor{blue}{sidelobe modulation} technique, amplitude modulation (AM), phase-shift keying (PSK) and quadrature AM (QAM) \textcolor{blue}{type of} modulations can be applicable for phased array radars. These modulations will give an opportunity to send only one information symbol per radar pulse. Recently, several researches are conducted to extend this method to \textcolor{blue}{MIMO} radars \cite{SignalingStrategiesDFRC}. To this end, simulations show that \textcolor{blue}{almost} megabits of information can be transferred \textcolor{blue}{by utilizing} MIMO radar waveform diversity. Fig. \ref{fig:DFRC} displays an example for sidelobe modulation based DFRC. 16-QAM symbols are generated at the \textcolor{blue}{sidelobe of the beampattern} with $16$ element transmit array.

\subsection{Radar Sensing Utilized by Communication Systems\textcolor{blue}{: Com-Rad}}

Mainly, OFDM-based communication systems drive the major trend of merging the radar sensing and communication researches, since OFDM waveforms can provide notable performance in radar sensing and imaging due to the use of multi-carrier waveforms\cite{IoTRadarCommJournal}. These systems use OFDM waveforms as a signal of opportunity for sensing operation. 

In order to reach richer radar performance with OFDM waveforms, researchers try to develop new methods to overcome major drawbacks introduced by information carried OFDM waveforms \cite{surveyJRC}. First, reliable \textcolor{blue}{radar} sensing cannot be guaranteed since the correlation properties of the OFDM waveform \textcolor{blue}{completely} depends on the information. Other drawback comes from high sidelobes at the range profile due to the auto-correlation function of the OFDM \textcolor{blue}{waveform}. To overcome these disadvantages, several OFDM radar processing techniques are proposed in the literature \cite{surveyJRC}. Although, generally fair radar performance can be achieved by OFDM-based communication systems when standardized waveforms are used, they can provide huge data bandwidth. Therefore, some researchers \textcolor{blue}{also} focus on OFDM like waveform designs which provide simultaneous operation.

\section{Possible Topologies for \textcolor{blue}{IoR}}

\begin{figure*}[t]
	\centering
	\includegraphics[width=0.8\textwidth]{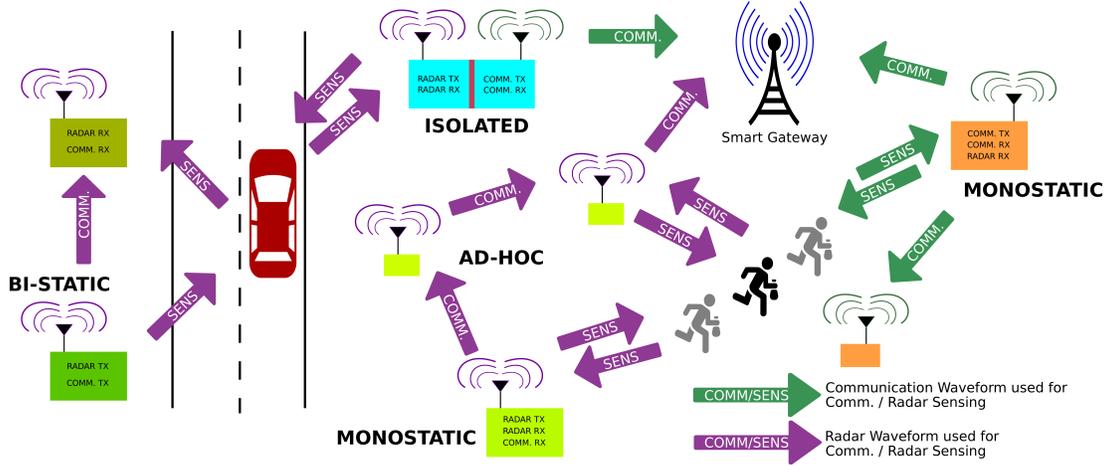}
	\caption{Possible topologies for the \textcolor{blue}{IoR}.}
	\label{fig:topology}
\end{figure*}

Various type, size and application specific radars can be found in the market for \textcolor{blue}{radar} sensing purposes. Most radar-equipped \textcolor{blue}{devices} require Internet to reach databases to provide services to customers. To establish connection to the Internet, devices can communicate \textcolor{blue}{with gateways or each others in} different topologies. \textcolor{blue}{These topologies are discussed in detail below, and illustrated in Fig.\ref{fig:topology}.} 

\subsection{Isolated}

\textcolor{blue}{Recent} solutions include separate communication and radar \textcolor{blue}{circuitries} which use the spectrum isolated manner via solving the interference mitigation \textcolor{blue}{issues}. Radars are completely detached from the communication infrastructure. Bluetooth, RFID, Wi-Fi, and telephonic data services are the most of the common wireless technologies responsible for carrying the radar related data to cloud services.

\subsection{Mono-static}

\textcolor{blue}{Most of the JRC technologies in the literature are proposed for the cases that transmitter and receiver circuitries are located on the same device, i.e. mono-static architecture. Radar coherent processing can be easily realized due to the lack of synchronization. This makes mono-static is the easiest topology to implement on IoR.}

\subsection{Bi-Static}

\textcolor{blue}{IoR topology can be counted as bi-static when the }transmitter and receiver \textcolor{blue}{are located} apart from each other with a meaningful distance. \textcolor{blue}{Bi-static topology may be required to broadcast to reduce communication or energy consumption by taking advantage of transmitter and receiver circuits as separate devices. Moreover, IoR in bi-static topology may offer multi-perspective information from the target.} However, high rate synchronization is required for bi-static operation. Although, there are several difficulties as synchronization, main benefit is that bi-static radar can inherently sense the channel without a need for any other mechanism. Therefore, this inherent estimation process definitely can support the communication equalization problems.

\subsection{Ad-Hoc}

This topology can be used when the devices have limited communication range or smart gateway is at relatively far distance. Wireless radar sensor network \textcolor{blue}{(WRSN) may} be counted as a major case of the ad-hoc topology. Radar nodes operate in a collaborative manner to enhance their own radar sensing capabilities. Each node has to conform the routing schemes to efficiently reach Internet via a sink node.

\section{\textcolor{blue}{Application Areas and Possible JRC technologies} for \textcolor{blue}{IoR}}
\label{Applications}

\begin{figure*}[t]
	\centering
	\includegraphics[width=0.9\textwidth]{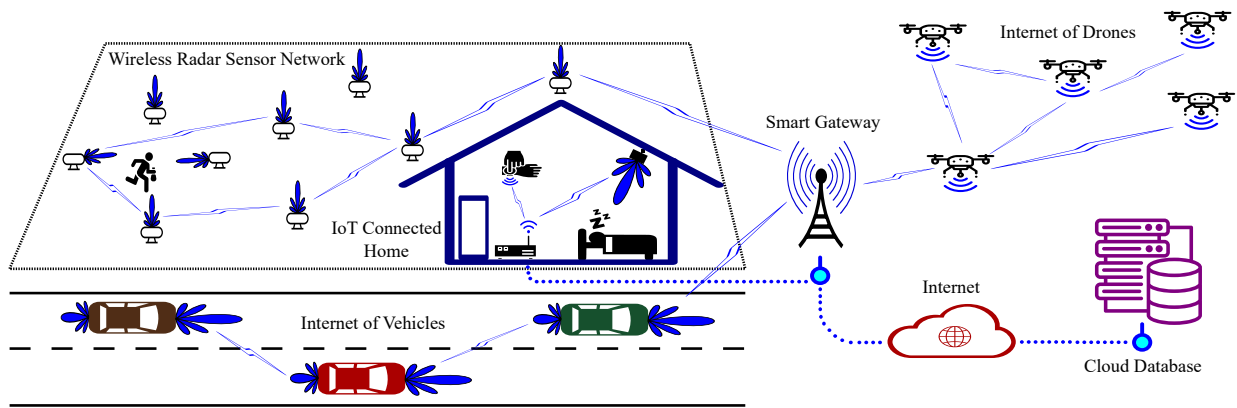}
	\caption{Illustration for the Internet of Radio Detectors And Rangers (RADARs).}
	\label{fig:jrcCoherentMIMO}
\end{figure*}

In this section, various application areas suitable for the concept of \textcolor{blue}{IoR} and \textcolor{blue}{possible JRC technologies} are discussed. These areas are selected as they are currently being researched in the literature or there are several radars equipped IoT products already in these areas. \textcolor{blue}{Fig. \ref{fig:jrcCoherentMIMO} illustrates the possible application areas for IoR.} Each of these applications has different requirements, limits and regulatory issues. Some of them already have their own communication standards, however, vast majority are at the beginning of the research. \textcolor{blue}{Table \ref{tab:specification} displays the possible JRC solutions with respect to the IoR application domains. For each application area, we have listed the specific features in terms of required data-rate, energy consumption, complexity, mobility and LoS status for the communication link. Then, applicable JRC solutions are explained in each discussion of the application area.}

\begin{table}[t]
	\caption{\textcolor{blue}{Possible JRC Solutions for IoR Application Areas}}
	\label{tab:specification}
	\begin{center}
		\resizebox{1\columnwidth}{!}{
			\begin{tabular}{|>{\color{blue}}c|>{\color{blue}}l|>{\color{blue}}l|}
				\hline\\[-1.1em]
				\emph{\textbf{Applications}} & \emph{\textbf{Features}} & \emph{\textbf{Possible JRC Solutions}} \\ \hline\\[-1.1em]
				\textbf{Personal} & Low power, low complexity &  \\ \\[-1.2em]
				(Wearables, & very short-range, high data-rate, & Coexistence (Isolated) \\ \\[-1.2em]
				smart-watch, etc.)& no Line-of-Sight & \\ \hline\\[-1.1em]
				\textbf{Home} & Low complexity, short-range, & Com-Rad \\ \\[-1.2em]
				(Tracking, presence, & no Line-of-Sight & (Monostatic, Multistatic) \\ \\[-1.2em]
				fall detection, etc.)&  & \\ \hline\\[-1.1em]
				\textbf{Medical} & Low complexity, short-range, &  \\ \\[-1.2em]
				(Respiration and & Line-of-Sight & Coexistence (Isolated) \\ \\[-1.2em]
				heartbeat rate, etc.)&  & \\ \hline\\[-1.1em]
				\textbf{Vehicles} & High data-rate, mobility,  & Co-design \\ \\[-1.2em]
				(Autonomous driving, & short/medium-range, & (Monostatic) \\ \\[-1.2em]
				safety, connected cars, etc.)& Line-of-Sight & \\ \hline\\[-1.1em]
				\textbf{Drones} & Low power, low complexity, & Com-Rad \\ \\[-1.2em]
				(Sense\&avoid, altimeter, & short/medium-range, mobility, & (Monostatic, Ad-Hoc) \\ \\[-1.2em]
				traffic management, etc.)& high data-rate, Line-of-Sight &  \\ \hline\\[-1.1em]
				\textbf{WRSN} & Low power, low complexity,  & Rad-Com \\ \\[-1.2em]
				(Intruder detection, & short-range, Line-of-Sight & (Monostatic, Ad-Hoc) \\ \\[-1.2em]
				tracking, imaging, etc.)&  & \\ \hline
			\end{tabular}
		}
	\end{center}
\end{table}

\subsection{Personal and Home}
\label{App:PersonalHome}

\textcolor{blue}{High amount of medical costs are spent over the years due to the non-fatal injuries in the house environment. Most injuries occur in older people's bathrooms. To keep them safe, monitoring in the bathroom may be required. However, placing cameras is not an option because of the privacy issues. Therefore, radar sensors can be a good candidate for monitoring people in their houses or schools without compromising their privacy. Hence, the radar can eliminate these privacy problems caused by the camera.}

Recent developments in chip technologies make the radar sensors possible to use in our houses as a surveillance sensor. These radar sensors can detect a person breathing and identify their gestures even if they are behind walls, smoke and steam \cite{RadarforAssistedLiving}. These systems require emitting radio waveform to operate. However, there are also passive radar applications works as a home alarm system. These passive radar sensors uses communication signals as an opportunity to detect movements in a home environment \cite{WifiPassiveRadarHome}.

\textcolor{blue}{Personal IoR devices are low power and compact gadgets and their sensing range is short. On the other hand, these sensors are located within a smart wearable device, such as smart-watch. Relatively high data-rate may be required for these wearables. \textcolor{blue}{Ultra-WideBand (UWB)} technology is preferred for sensing due to its very low power consumption, high range resolution and accuracy. To classify gestures coming from very small movements as in the example of smart-watch with finger gesture control, emission power is mostly limited to some level. Therefore, establishing a link with high data-rate seems to be not feasible. In terms of current JRC technologies, we have suggested using dual circuitry, i.e. isolated topology, for this type of application area. However, coexistence based JRC techniques can still be applied to reach advance sensing and communication missions.}

\textcolor{blue}{Most houses have wifi capable devices as routers to reach Internet. For home sensing applications, wifi signals already proposed in the literature for tracking, presence sensing and fall detecting purposes. Com-Rad JRC techniques can be applied with some modifications to enhance sensing performance.}

\subsection{Medical and Health-care}
\label{App:Medical}

Recent studies reveal that radar \textcolor{blue}{based} medical devices can be helpful in medical imaging for tracking \textcolor{blue}{a} tumor during radiotherapy. \textcolor{blue}{Radars} can be used for measuring vital signs remotely \cite{NonContactHealtcareMonitoring}. These radars are capable of detecting the movements of human chest of a \textcolor{blue}{stationary} person caused by cardiopulmonary activities. Respiration and heartbeat rates can be estimated non-contact manner without connecting any devices. Most of the solutions \textcolor{blue}{utilizes} UWB \textcolor{blue}{technology} to achieve millimeter resolution. UWB technology helps the radar to meet low power requirements as IoT sensor. Using these sensors in our \textcolor{blue}{home} is a very delicate issue. \textcolor{blue}{JRC technology} can be a great solution for keeping transmit power lower than the human safety limits while achieving meter-level radar detection range.

\textcolor{blue}{Current heartbeat and respiration rate capable medical IoR devices can measure very small movements under any clothing using the benefits of UWB technology. Although the primary mission of this IoR device is the radar sensing, single emission may not be applicable due to the radar antenna directivity and power levels. Moreover, short-range directional UWB emission may prevent establishing a link. Coexistence JRC techniques are suggested for medical IoR devices with dual-circuity as in Table \ref{tab:specification}. }

\subsection{Vehicles}
\label{App:Transportation}

Last decades, IoT technology reveals itself in every vehicle. Vehicle-to-vehicle (V2V) communication technology is slightly replaced by first vehicle-to-everything (V2X) communications. Now, internet-of-vehicles (IoV) is defined as communication of vehicles, infrastructures, grids and hand-held devices carried by pedestrians using the V2X technology. Up to now, only two communication standards are proposed for V2X communications, as IEEE-802.11p (DSRC) and 3GPP (C-V2X).

For autonomous vehicles, Lidar, automotive radars and visual cameras are the most important sensors to ensure safety. Visual cameras generate faulty results when sunlight hits its CMOS sensor. Equipping systems with Lidar is the best solution for extracting 3D mapping \textcolor{blue}{to} classifying object around the vehicle. However, Lidar systems can not discriminate lane markings under heavy weather conditions. Radar will become a key player for meeting the safety requirements from ISO 26262 under all weather conditions. 

Several types of automotive radars \textcolor{blue}{are} in the market. Mid and long range one's help to providing necessary information for adaptive cruise control, emergency breaking and highly automated highway driving type operations. Ultra short and short range ones are used for parking, cross-traffic alerts, lane change assistance, blind spot detection and collision avoidance systems. Also, gesture control, occupant detection and driver monitoring applications can be done by proximity radar sensors. These applications require exchanging considerable amount of data in a second generated by radars vary from kilo-bits to hundreds of mega-bits. 

Up to now, communication waveform detailed in IEEE 802.16p and IEEE 802.16ad standards are exploited for radar purposes. Researchers propose new ways of using V2V \textcolor{blue}{waveform} as an opportunity to detect and track targets, and also form radar images. Due to the standardization of V2X, C-band with maximum 80MHz bandwidth is reserved. Hence, when considering more and more sensors in a single vehicle, more bandwidth \textcolor{blue}{will} be required in near future. Higher bands, which are allocated for automotive radars, seem suitable to provide notable bandwidth for V2X. Currently, researchers are spending more effort to find techniques in millimeter bands for supporting JRC \cite{mmWaveVehicularCommunicationSensing}. \textcolor{blue}{Due to the high data-rate and mobility requirements for V2X, Com-Rad JRC techniques can be selected for the vehicles in IoR. However, achieving the best radar imaging and tracking performance are also important for the safety reasons. Therefore, co-design solutions appear as the best option for the IoR devices in vehicles. Moreover, new type of medium access schemes must be developed to reach high data-rate and sensing performance.}

\subsection{Drones}
\label{App:Drones}

Nowadays, drones have attracted a lot of attention due to their ability to visit areas that are difficult to reach by traditional methods. In the near future, millions of drones are predicted to be separated around the world. There are many applications that the drones are used. Package delivery, border security, traffic and wild life surveillance, search and rescue, agriculture, and entertainment can be counted as the part of major drone applications. 

Due to its worldwide usability, researchers begin to work on managing the airspace and coordinating aerial vehicles through technologies as ADS-B and V2V communication. At this point, Internet of Drones (IoD) concept is proposed \cite{IoDrones}. IoD is a layered network control architecture designed for regulating airspace with providing location services. Despite all of these efforts, on-board collision avoidance sensors are still needed to \textcolor{blue}{ensure} safety. Due to its performance under heavy weather, radar is the big candidate for being the collision avoidance sensor on drones. Researchers try to find new ways of developing compact and energy efficient radars to deploy them even on a small UAV. Not only as a collision avoidance sensor, radars are studied for several purposes as surveillance, imaging and altimeter \cite{RadarTakingOff}.

Energy efficiency is an important aspect for drones phenomenon. Instead of carrying dual RF equipment, combining radar and communication in a single radio circuitry will be a great solution to conserve more power. Therefore, \textcolor{blue}{JRC} solutions should be investigated when determining communication standards and regulations for IoD applications.

\textcolor{blue}{In near future, sense\&avoid and altimeter missions can be covered by next generation communication techniques, i.e. 5G NR. Same as vehicles, Com-Rad JRC techniques appear as the best option for the IoR devices in drones, due to the high data-rate and mobility requirements.}

\subsection{Wireless Radar Sensor Networks}
\label{App:Security}

Although, IoT does not dedicate a specific communication infrastructure, wireless technologies constitute a large part of it. Especially, developments in Bluetooth technology play an important role in the growth of the IoT products market. The small, cheap and low powered wireless sensor network (WSN) sensors will bring a new vision to the IoT domain for deploying even the smallest devices in any kind of environment. Integrating WSN into IoT cloud will be a significant evolution of WSNs.

Low power and small scale radar sensors are introduced into the market with the trust of the \textcolor{blue}{UWB} technology. Then, radar sensor becomes the noticeable part of the wireless sensor network structures. This specific type of WSN architecture generally called as wireless radar sensor network (WRSN). Using these networked structure of radar sensors, precise localization, tracking or even imaging of moving objects in harsh propagation environments can be possible. 

Energy efficiency plays a major role in IoR applications. Small and low power WRSN nodes could not find sufficient energy for realizing dual role of operation as communication and radar sensing with separate circuitries. Therefore, joint radar-communication techniques need to be coupled into WRSN structures.

\textcolor{blue}{IoR devices in WRSN have low computational power and limited battery capacity. Their sensing range is very short and they have to be communicate with sink via ad-hoc manner. Applying Rad-Com JRC techniques appear as the best option for the WRSN devices due to low data-rate, ad-hoc topology and limited sensing requirements.}

\section{\textcolor{blue}{Challenges for IoR}}

\textcolor{blue}{Most of the IoT products in the market have some limitations due to their application scenarios. These limitations bring several challenges while realizing \textcolor{blue}{IoR}. This section discusses the challenges and overcoming techniques to realize IoR.}

\subsection{\textcolor{blue}{Complexity}}

\textcolor{blue}{Depending on the applications, IoR devices mainly have small and compact design, and limited battery and computational power. Operating the advanced JRC techniques almost impossible to implement in a compact IoR device. Due to these concerns, JRC techniques has to be selected carefully with respect to application area and its hardware complexity.}

\textcolor{blue}{Rising trend on intent to use the mmWave spectrum opens a vision of developing radar on chip (RoC) applications. These RoC devices has very small footage, however they have enough space to implement $192$ virtual MIMO array radar. Also, these RoC devices have two CPUs and two DSPs, and a capability of up to $20$ Tera-Ops baseband processing \cite{UhnderMIMO}. Hence, in near future, even complex JRC algorithms can be implemented this type of system on chip solutions within a compact package.}

\subsection{Adaptivity}

\textcolor{blue}{Providing dual-mission capability to an IoT device opens a way to reduce costs and manage the spectrum more efficiently. However, this progress arises with the new problem of balancing the needs of the two different disciplines without any interference. Trade-off between sensing or sending has to be parameterized to cope with different IoR application scenarios and environments. Special multi-access channel methodologies can be utilized for IoR where the incorporated targets are accepted as virtual users. These methodologies must present more reliable radar and communication performance while preserving more energy. }

\subsection{Energy Consumption}

With IoT revolution, every device \textcolor{blue}{can} become smart and connected. Connectivity requires more energy and \textcolor{blue}{more} spectrum. This makes energy consumption be the one of the most critical issue for the IoT devices.\textcolor{blue}{ To enable IoR concept, applicable JRC techniques and medium access schemes have to be deliberately developed to draw less energy.}

\begin{figure}[t]
	\centering
	\includegraphics[width=1\columnwidth]{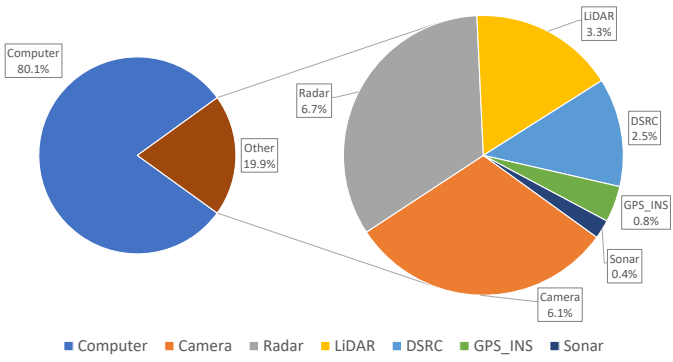}
	\caption{Power consumption by subsystems of a Medium CAV \cite{LifeCycleAutomatedVehicles}.}
	\label{fig:PowerMediumCAV}
\end{figure}

IoT technology also provides a communication infrastructure for autonomous vehicles. The connected and automated vehicle (CAV) is an important and emerging technology that could totally transform the current transportation system vision. The recent study in \cite{LifeCycleAutomatedVehicles} \textcolor{blue}{presents} the power consumption for a medium level CAV system in Fig.\ref{fig:PowerMediumCAV}. \textcolor{blue}{Then}, \textcolor{blue}{maximum percentage of} the power is consumed by the main computer that performs sensor signal processing operation\textcolor{blue}{s}. With the developing \textcolor{blue}{chip} technology, this consumption will be drawn to a level that will not be problematic. However, deployed sensors always be the important part of the CAV systems, and surely, their power consumption have to be limited to a certain level. Together with radar and DSRC system consume almost the half of the total power consumption of the sensor systems. In the future, these RF systems have to merge to drawn less energy. Further, next generation JRC \textcolor{blue}{technologies must ensure} low power consumption without any connectivity and safety problems.

\section{Future Perspectives}

\subsection{Architecture for \textcolor{blue}{IoR}}

There is no general agreement on layered architecture for IoT. Common IoT architecture is considered to have three layers which are the application, network and perception layer. Then, another layers were added to the list and IoT architecture becomes five layered structure. Recently, CISCO proposed a seven layered architecture which exploits processing and transport layer in five layer architecture.

We mainly focus on perception and transport (or connectivity) layer for realizing \textcolor{blue}{IoR} architecture. In traditional radar equipped IoT \textcolor{blue}{devices}, radar sensing is \textcolor{blue}{occurred} in perception layer and communication operation is held in transport layer. JRC technology \textcolor{blue}{enables} perception and transport operations done in single emission. Therefore, all the control in two layer has to be done in a bounded manner. The performance of radar sensing intuitively limited to the communication performance and vice versa. As \textcolor{blue}{a} basic example, radar systems work on \textcolor{blue}{the} principle of transmit and receive cycles characterized by pulse repetition frequency (PRF). In JRC systems, PRF mainly drives the data-rate, and \textcolor{blue}{higher} PRF results increased data-rate. Under good channel conditions, higher data-rates may be desired in IoR applications. On the other side, radar range and \textcolor{blue}{tracking} performance are affected by the PRF. To reach \textcolor{blue}{the} best operation, this trade-off needs to be \textcolor{blue}{handled} by some control mechanism which directly access the transport and perception layer. We called this layer as ``interoperability layer". Some calculations must be done under this layer to fully optimize radar sensing and communication operation.

\subsection{Socially Cognitive Radar}

\begin{figure}[b]
	\centering
	\includegraphics[width=0.75\columnwidth]{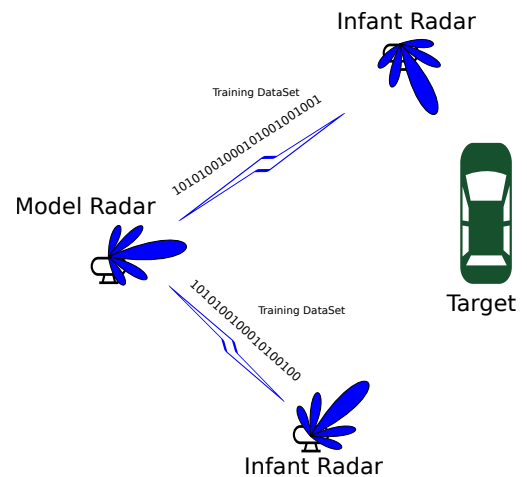}
	\caption{Observational Learning in \textcolor{blue}{IoR}}
	\label{fig:ObservationalLearningIoRADARs}
\end{figure}

\textcolor{blue}{Current} radar technology is going toward to be more smart systems due to the recent advances in cognitive radar technologies. In near future, it is probable to have cognition capable IoR devices in the market. In \textcolor{blue}{recent} radars, receiver is usually adopted to the radar environment. For a cognitive operation, adaptivity has to be extended to the transmitter side too. In addition, radars should use their knowledge to overcome problems caused by \textcolor{blue}{agile} targets. However, upgrading radars with non-centric cognition will also bring several problems. With cognition, we grant an ability of making decision by itself to radar in each cognitive cycle. However, with no central management, this may end up with a chaos \textcolor{blue}{in a network}. Each cognitive radar may select same spectrum or waveform to reach maximum operational performance, thus this leads to congestion on the radar mission. To prevent this congestion, JRC techniques can be utilized. 

Social skills that we have, help us to keep society away from anarchy. We can inspire from the human social behaviors with \textcolor{blue}{inspiring the} Social Cognitive Theory (SCT) to prevent chaos in network. SCT is the view that people learn by observing others. This theory proposes that parts of a person's knowledge achievements can be totally correlated to watching others via social experiences and corporations. \textcolor{blue}{Albert} Bandura mentioned that when a person notices other individual else rewarded due to his behavior, he/she has an intention to behave in the same way to reach an award. According to Bandura, there are five major notions related with the SCT framework. These \textcolor{blue}{notions} can be listed as observational learning, outcome expectations, self-efficacy, goal setting and self-regulation \cite{Bandura:01}.

In a cognitive radar network, each social radar is ready to observe its own signals reflected from the environment, however, merging observation of other radar’s signals or communication mechanism into cognitive radar learning process takes the cognitive radar one step further to be a social cognitive radar. This can be realized via a special communication network or enhanced radar signal processing algorithms. The output of this mechanism should be converted to a training data set at the model radar and it has to be an input to the infant radar cognitive memory \textcolor{blue}{as shown} in Fig.\ref{fig:ObservationalLearningIoRADARs}. 

One of the main realization tool for observational learning is establishing communication links between \textcolor{blue}{cognitive radars}. \textcolor{blue}{They} require a global feedback to do its job. In network architecture, each radar needs a link to establish this global feedback mechanism. Especially, for a cognition capable IoR devices with distributed cognition, each link has to be complied with strict timing limitations. In this nature, a need for a new type of communication network architecture arises. In the literature this type of network is yet to be developed for cognition capable \textcolor{blue}{IoR devices}.

\section{Conclusion}
\label{con}
Nowadays, developing single emission dual-purpose technologies have attracted a lot of attention due to their capability of eliminating congested spectrum problems. Joint radar-communication technology becomes the major research field for this purpose. Its compactness and energy efficiency features make JRC technology a promising solution for IoT platforms. However, single emission emerges as an optimization problem between radar sensing and information sending. This problem area spurs the creation of a new concept for the Internet of RADARs (IoR). This article is an attempt to present a forward-looking research direction for developing architectures to enable \textcolor{blue}{IoR}. While a partial vision was sketched, we hope our discussion will elevate interests on the future evolution of IoT habitat.

\textcolor{blue}{
\begin{IEEEbiography}{Muharrem Arik} received his Ph.D. in Electrical and Electronics Engineering from Koc University, Istanbul, Turkey, in 2019. He is currently Radar Test Lead in Radar and Electronic Warfare Systems Business Sector in ASELSAN Inc. His current research interests include joint radar-communications, cognitive radar networks, cognitive electronic warfare. He received the Thesis of the Year Award 2010 given by Middle East Technical University Graduate School of Natural and Applied Sciences.
\end{IEEEbiography}}

\begin{IEEEbiography}{Ozgur B. Akan [M’00, SM’07, F’16]} received his Ph.D. degree in electrical and computer engineering from the Broadband and Wireless Networking Laboratory, School of Electrical and Computer Engineering, Georgia Institute of Technology, Atlanta in 2004. He is currently with the Electrical Engineering Division, Department of Engineering, University of Cambridge, United Kingdom, and also the director of the Next-Generation and Wireless Communications Laboratory in the Department of Electrical and Electronics Engineering, Koc University. His research interests include wireless, nano, and molecular communications, and the Internet of Everything.
\end{IEEEbiography}

\end{document}